\documentclass[12pt]{article}

\usepackage{amsmath,amssymb}

\title{Hilbert's ``World Equations" and His Vision of a
Unified Science\footnote{To appear in a forthcoming volume of {\it
Einstein Studies}, eds. J.~Eisenstaedt and A.J.~Kox.}}

\author{U.~Majer\thanks{Hilbert-Edition,
Universit\"at G\"ottingen, Papendiek 16, D-37073 G\"ottingen,
Germany, umajer@gwdg.de} and T.~Sauer\thanks{Einstein Papers
Project, California Institute of Technology 20-7, 1200
E.~California Blvd., Pasadena, CA~91125, USA,
tilman@einstein.caltech.edu}}

\date{}

\begin{document}

\maketitle

\begin{abstract}
In summer 1923, a year after his lectures on the `New Foundation
of Mathematics' and half a year before the republication of his
two notes on the `Foundations of Physics,' Hilbert delivered a
trilogy of lectures in Hamburg. In these lectures, Hilbert
expounds in an unusually explicit manner his epistemological
perspective on science as a subdiscipline of an all embracing
science of mathematics. The starting point of Hilbert's
considerations is the claim that the class of gravitational and
electromagnetic field equations implied by his original
variational formulation of 1915 provides valid candidate `world
equations,' even in view of attempts at unified field theories \'a
la Weyl and Eddington based on the concept of the affine
connection. We give a discussion of Hilbert's lectures and, in
particular, examine his claim that Einstein in his 1923 papers on
affine unified field theory only arrived at Hilbert's original
1915 theory. We also briefly comment on Hilbert's philosophical
viewpoints expressed in these lectures.
\end{abstract}

\section{Introduction}

In the history of unified field theory, many contributors may be
identified \cite{GoldsteinRitter2003,Goenner2004}, among them
certainly, and perhaps foremost, Einstein. Hilbert's place in the
history of the unified field theory program is also well
recognized (see, e.g., the discussion of his work in Vizgin's
study \cite{Vizgin1994}). But we tend to view the history of
physics in which Einstein was involved through that scholarship
which has focussed exclusively, or at least predominantly, on
Einstein's work as such. For the case of Einstein's ``later
journey,'' we believe that many physicists as well as historians
would subscribe to Pais's verdict that ``his work on unification
was probably all in vain'' \cite[p.~329]{Pais1982}. The dismissal
of Einstein's efforts over three decades is to some extent
supported by Einstein's own self-image, in his later years, as the
``lonesome outsider'' working without real appreciation in his
golden Princeton cage. Einstein was an original thinker and an
influential voice in the debate, and for this reason understanding
Einstein's obsession with the problem of a unified field theory
over the last thirty years of his life presents as much of a
challenge to the historian as understanding the achievements of
his early work.

To this purpose, it helps to free one's mind from preconceptions.
We then find Hilbert's insights of great advantage since he was
both knowledgeable and had a well-founded and original perspective
of his own.

Let us make a distinction right at the beginning in order to
disentangle different scientific approaches. The problem of a
unified field theory, as of the 1920s, can be seen in a more
specific sense as the problem of finding a consistent and
satisfactory mathematical unification of the gravitational and
electromagnetic fields, be it by modified field equations, by a
modification of the space-time geometry, or by increasing the
number of space-time dimensions. But there is another aspect to
the problem that is, we believe, of both historical and
philosophical interest. This aspect concerns the way in which
contemporary scientists perceived the technical problem of
unification in the wider context of a unified corpus of human
knowledge and understanding. In this respect, Hilbert's
perspective on the mathematical sciences as an integrated whole
can contribute to our modern attempts to come to grips with the
philosophical implications of an ever increasing specialization in
the natural sciences. Hilbert certainly was not the only one who
envisaged a unified science at the time. Many contemporary
mathematicians shared this concern. Felix Klein's {\it History of
the Development of Mathematics in the 19th century}
\cite{Klein1979} can also be seen as a most interesting attempt to
understand the inner organic unity of the corpus of mathematical
knowledge. Other names that come to mind immediately are those of
Kaluza and Weyl, but the list certainly does not end here.

Einstein, too, shared this interest in understanding the inner
unity of our knowledge of nature, and for him, too, the problem of
finding a mathematical representation that would provide a
unification of the gravitational and electromagnetic fields was
more than just a technical problem. This aspect of his work is
expressed most convincingly in Einstein's own account of his
lifelong research concerns as given in his 1949 {\it
Autobiographical Notes} \cite{Einstein1949}. Einstein, as we will
argue, followed in his later work a path that is not at all very
different from Hilbert's. Hilbert himself perceived Einstein as
sharing his concern. Of course, there are differences, which we do
not deny. But from a broader perspective, both Einstein and
Hilbert –-- and others, one may add –-- belong to a tradition
which attempts to integrate our human knowledge and to perceive an
inner unity in science. For today's philosophers, this tradition
seems to belong to the 18th and 19th century rather than to the
20th, or to the 21st, for that matter. In this respect, Einstein
and Hilbert are akin more to the encyclopedists and enlightenment
natural philosophers than to modern puzzle solvers.

\section{Hilbert's Lectures on Fundamental Questions of Modern Physics of
1923}

The document to which we would like to draw attention in this
paper is a manuscript extant in the Hilbert archives in
G{\"o}ttingen. It will be published in one of the physics volumes
of the Hilbert Edition under the title ``Fundamental Questions of
Modern Physics.'' It is a batch of roughly 100 manuscript pages
with notes for a trilogy of lectures that Hilbert delivered at the
end of the summer semester of 1923 in Hamburg.\footnote{The
lectures were held in Hamburg on July 26, 27, and 28, 1923. They
were announced under the title ``Grunds\"atzliche Fragen der
modernen Physik,'' see ``Hamburgische Universit\"at. Verzeichnis
der Vorlesungen. Sommersemester 1923,'' Hamburg 1923, p.~41. The
third of the three lectures was delivered a second time, with
short summaries of the first two lectures, in a lecture held at
the ``Physikalische Gesellschaft'' in Z\"urich on October 27,
1923. This lecture was announced under the title
``Erkenntnistheoretische Grundfragen der Physik,'' see ``Neue
Z\"urcher Zeitung,'' Nr.~1473, Erstes Morgenblatt, 27 October,
1923. The manuscript Cod.~Ms.~Hilbert 596 in the {\it
Handschriftenabteilung} at the {\it Nieders\"achsische Staats- und
Universit\"atsbibiliothek} (NSUB) contains the notes for both the
Hamburg and Zurich lectures. It will be cited in the following as
{\it Lectures}.} The three lectures focus on three different
topics: the first deals with what Hilbert called the ``World
Equations,'' where these equations are introduced; the second part
discusses applications and consequences of those equations; and
the third lecture contains a discussion of the old problem of
theory and experience.

To Hilbert at that time, the epistemological and philosophical
implications of recent developments in physics were of central
concern. He himself had contributed substantially to modern
mathematical physics in the preceding years, most notably through
his two Communications to the G{\"o}ttingen Academy Proceedings on
the "Foundations of Physics" of November 1915 and December 1916,
respectively \cite{Hilbert1915,Hilbert1917}. By the summer of 1917
at the latest, however, another problem was increasingly occupying
Hilbert's mind, namely the problem of an absolute consistency
proof of arithmetic that would provide a sound logical foundation
for the whole body of mathematics. Just as in Hilbert's work in
physics, the roots of this preoccupation date back to his very
early work, at least to his ``Mathematical Problems" of 1900
\cite{Hilbert1900}. This interest resurfaced with a lecture on set
theory held in the summer term of 1917.

As a matter of fact, Hilbert's renewed attention to the
foundations of mathematics in general, and to a theory of proof in
particular, contributed to his taking a broader perspective on the
contemporary debates in General Relativity and Field Theory. He
had kept an active interest in the development of General
Relativity after 1915 but was increasingly concerned with the
philosophical implications of the new theories rather than with
contributing solutions of some of its outstanding technical
problems.\footnote{In this respect, we disagree with the claim
made by Renn and Stachel, who characterize Hilbert's work in GRT
as the transition from a ``Theory of Everything to a Constituent
of General Relativity,'' \cite{RennStachel1999}. While their
assessment may be true in abstraction of its actors, it is
certainly not true for Hilbert himself. Rather than beginning to
see his own work as a constituent of General Relativity, his main
effort with respect to General Relativity in later years was to
emphasize his claim that his approach would provide the basis for
a true unification of physics.} He also began to spend a great
deal of energy in popularizing these new developments and in
acquainting a larger audience with the results of modern physics.
It is therefore no accident that when Hilbert spoke on the same
topic a few weeks later in Zurich, but in a single lecture, he
chose to center on the third of his Hamburg lectures.\footnote{The
lecture was arranged by Peter Debye following Hilbert's request:
``Prof.\ Hilbert who is presently staying in Switzerland wished to
deliver a lecture in the joint Physical and Mathematical
Colloquium.'' (``Herr Prof.~Hilbert, welcher zur Zeit in der
Schweiz weilt, hatte den Wunsch im zusammengefassten Physik. und
Mathematischen Kolloquium einen Vortrag zu halten.'') P.~Debye to
Robert Gnehm, 22 October 1923. Archiv des Schweizerischen
Schulrats, ETH-Bibliothek, Z\"urich.} He used the same manuscript
notes for the Zurich lecture, but since he had to cut down the
material, he summarized the main points of the first two lectures.
This editing of his own manuscript makes it difficult to exactly
associate specific phrases with either the Hamburg or Zurich
lectures.

\section{Hilbert's ``World Equations'' of summer and
fall of 1923}

Hilbert starts his first lecture by introducing what he calls the
``World Equations" or the ``World Laws" (``Weltgleichungen" or
``Weltgesetze"). The way Hilbert introduces these equations is
interesting in itself but for the sake of brevity, we shall only
say that these equations basically are the same ones that he had
proposed in his First Communication on the Foundations of Physics
\cite{Hilbert1915}, considering the fact that Hilbert had,
originally, not completely specified the Lagrangian $I$ of the
variational integral
\begin{equation}
\int I \sqrt{-g}d\tau,
\end{equation}
where $g=\operatorname{det}(g_{\mu\nu})$ and the integral is over
(a domain of) four-dimensional space-time. But both in 1915 and
now again in 1923 he pointed out that the fundamental dynamical
variables are the ten components $g_{\mu\nu}$ of the metric tensor
and the four components $\varphi_{l}$ of the electromagnetic
potential.%
\footnote{As an aside, Hilbert observed in his 1923 lectures that
the difference between his own fundamental equations of November
1915 and Einstein's gravitational field equations pertains to the
choice of fundamental variables: ``Einstein's equations of
gravitation are, in the sense defined here, the fundamental
equations of physics, if one takes in them the gravitational
potential $g_{\mu\nu}$ and the energy tensor as fundamental
potentials. I proposed, at the same time, fundamental equations of
physics, in which only the electromagnetic four-potential
$\varphi_{\rm k}$ enters in addition to the gravitational
potential $g_{\mu\nu}$.'' (``Die Einsteinschen
Gravitationsgl[eichungen] sind in dem hier definirten Sinne die
Grundgl[eichungen] der Physik, wenn man darin das
Gravitationspotential $g_{\mu\nu}$ und ausserdem den Energietensor
als Grundpotentiale nimmt. Ich habe zurselben Zeit
Grundgl[eichungen] der Physik aufgestellt, in denen neben dem
Gravitationspotential $g_{\mu\nu}$ nur noch das elektromagnetische
Viererpotential $\varphi_{\rm k}$ als Grundpotential auftritt.'')
{\it Lectures}, part I, p.~16.}

In his Hamburg and Zurich lectures, he takes the Lagrangian to be
the sum of a gravitational part $K$ and a matter part $L$,
\begin{equation}
I = K + L.
\end{equation}
The gravitational part $K$ is understood to be the Riemann
curvature scalar. The matter part $L$ is taken to be a sum of a
term proportional to the square of the fields, and another term
proportional to the square of the potential,
\begin{equation}
L = \alpha \Phi + \beta \varphi,
\label{eq:Lagrangian}
\end{equation}
where $\Phi \equiv \sum \Phi_{\rm mn}\Phi^{\rm mn}$ with
$\Phi_{\rm mn} \equiv \varphi_{\rm [m;n]}$\footnote{We are closely
following Hilbert's and Einstein's notation, with the following
exceptions: for notational brevity, we denote partial (coordinate)
derivatives by a subscript index separated by a semicolon (comma),
and indicate (anti)symmetrization by setting the relevant indices
in (square) brackets. We also do not use an imaginary
$x_4$-coordinate, as Hilbert did.} denoting the electromagnetic
field, and $\varphi \equiv \varphi_{\rm
k}\varphi^{\rm k}$.%
\footnote{Already in his First Note on the Foundations of Physics
\cite{Hilbert1915}, Hilbert had left open the final choice of a
matter term in the Lagrangian. It should be diffeomorphism
invariant, and it should not depend on the derivatives of the
metric. But Mie's example of a term proportional to the sixth
power of $\varphi$ had obviously been unacceptable, and a
different specification of the Lagrangian that would allow for
solutions of a reasonable physical interpretation had not yet been
found, see \cite{Mie1912} and also the discussion in Hilbert's own
lecture notes on ``Die Grundlagen der Physik,'' of the summer of
1916, which are located at the library of the Mathematics
Institute of G\"ottingen University, see especially \S\S~27--30.
For further discussion of Hilbert's First Communication, see
\cite{Sauer1999}.} As usual, variation with respect to the
components of the metric tensor produces the gravitational field
equations,
\begin{equation}
K_{\mu\nu} = -\frac{\partial \sqrt{-g}L}{\partial g^{\mu\nu}},
\label{eq:grav}
\end{equation}
and variation with respect to the components of the
electromagnetic four-potential produces generalized Maxwell
equations of the form
\begin{equation}
{\cal D}iv \Phi^{\rm mn} = \frac{\beta}{\alpha}\varphi^{m}.
\label{eq:elm}
\end{equation}
The latter equations are determined by the matter term alone. More
specifically, the first term in (\ref{eq:Lagrangian}) produces the
left hand side of the inhomogeneous Maxwell equations,
$\alpha{\cal D}iv \Phi^{\rm mn}$, while the second term in
(\ref{eq:Lagrangian}) produces a term proportional to the
electromagnetic vector potential, $\varphi^k$, the latter acting
as the source of the inhomogeneous Maxwell equations. Following
Mie's approach, external currents and charges are not part of the
theory. The homogeneous field equations,
\begin{equation}
\Phi_{\rm (mn;k)} = 0,
\label{eq:homMaxwell}
\end{equation}
follow, in the usual way, from the definition of the field and the
fact that the connection was assumed to be the symmetric
Levi-Civita connection.

\section{Hilbert's Comments on Einstein's Recent Work on
         Affine Field Theory}

At this point, Hilbert introduces a remark which at first sight
may seem preposterous, or, if you wish, arrogant and self-serving.
He claims that Einstein, in his most recent publications, would
have arrived, after ``a colossal detour,'' (``kolossaler Umweg'')
at the very same results and equations that Hilbert had put
forward in his first note on the Foundation of Physics of November
1915. But before dismissing this claim as a stubborn and senile
insistence of a mathematician who ``has left reality behind" let
us examine his claim more closely and see whether it is conducive
to a more nuanced historical interpretation.

The starting point is Hilbert's claim that the invariance of the
action integral allows one to interpret the electromagnetic field
equations as implicit in the gravitational field equations.
Hilbert here reiterates the claim of his first note that this fact
would provide the solution to a problem that he traces back to
Riemann, namely the problem of the connection between gravitation
and light. He goes on to observe that since then many
investigators had tried to arrive at a deeper understanding of
this connection by merging the gravitational and electromagnetic
potentials into a unity. The one example Hilbert mentions
explicitly is Weyl's unification of the two fields in a ``unified
world metric," as he calls it, by means of Weyl's notion of gauge
invariance. Another approach would be Eddington's who proceeded by
selecting ``certain invariant combinations'' as fundamental
potentials of the quantities determining the fields. Schouten then
had investigated the manifold of possibilities of such
combinations and realized that there would be a rich variety of
them. At this point, Hilbert inserts his comment on Einstein's
recent work. He says explicitly:
\begin{quote}
Einstein finally ties up to Eddington in his most recent
publications and, just as Weyl did, arrives at a system of very
coherent mathematical construction.
\end{quote}
But, Hilbert goes on,
\begin{quote}
However, the final result of Einstein's latest work amounts to a
Hamiltonian principle that is similar to the one that I had
originally proposed. Indeed, it might be the case that the content
of this latest Einsteinian theory is {\it completely equivalent}
to the theory originally advanced by myself.%
\footnote{``Einstein endlich kn\"upft in seinen letzten
Publikationen an Eddington an und gelangt ebenso wie Weyl zu einem
mathematisch sehr einheitlich aufgebauten System. Indess m\"undet
das Schlussresultat der letzten Einsteinschen Untersuchung wieder
auf ein Hamiltonsches Prinzip, das dem urspr\"unglich von mir
aufgestellten gleicht; ja es k\"onnte sein, dass diese
Einsteinsche Theorie inhaltlich sich mit der von mir
urspr\"unglich aufgestellten Theorie {\it v\"ollig deckt}.'' {\it
Lectures}, part~I, p.~19 (Hilbert's emphasis).}
\end{quote}
It is important to note that Hilbert makes his claim somewhat more
specific than that. Looking at the variational principle which he
explicitly writes down in the form
\begin{equation}
\delta\int\int\int\int \Big\{K+\alpha\Phi + \beta\varphi \Big\}
\sqrt{-g} dx_1dx_2dx_3dx_4 = 0, \label{eq:varprinc1}
\end{equation}
he observes that Einstein in his latest note had arrived at the
very same Hamiltonian principle
\begin{quote}
where $\varphi$ is defined through $\varphi^{\rm m} =
\operatorname{Div}\Phi^{\rm mn}$ and variation with respect to
$g_{\mu\nu}$ and $\Phi^{\rm mn}$ produces the eqs.\ $\Phi_{\rm
mn}=\operatorname{Rot}\varphi_{\rm m}$ instead of my
eq.~[(\ref{eq:elm})].
\end{quote}
Hilbert concludes:
\begin{quote}
Hence, nothing else than {\it an exchange of the two series}
[of] Max[well] eq[uations].%
\footnote{``... wo $\varphi$ durch $\varphi^{\rm m} =
\operatorname{Div}\Phi^{\rm mn}$ definiert ist und durch Variation
nach $g_{\mu\nu}$ und $\Phi^{\rm mn}$ die Gl.\ $\Phi_{\rm
mn}=\operatorname{Rot}\varphi_{\rm m}$ an Stelle meiner
Gl.~(\ref{eq:elm}) entstehen. Also Nichts als eine {\it
Vertauschung der beiden Serien} [von] Max.\ Gl.'' ibid., p.20
(Hilbert's emphasis).}
\end{quote}
The emphasis in the last quote is Hilbert's. He was not only
pointing at a vague similarity between his own work and
Einstein's. Rather he had identified the differences in their work
as being of a purely nominal nature.

\section{Einstein's ``colossal detour"}

In view of this remark, let us briefly examine Einstein's
post-1915 work in General Relativity, in particular with regard to
the problem of unifying gravitation and electromagnetism (see also
\cite{Vizgin1994,Goenner2004}).

Until 1923, it is perhaps not too unjust to say that Einstein
basically had been reacting to the work of others. He had
submitted Kaluza's theory of a five-dimensional metric for
publication in the Prussian Academy Proceedings \cite{Kaluza1921}
and had himself done calculations along this approach, partly in
collaboration with Jakob Grommer \cite{EinsteinGrommer1923}. He
had also published a couple of notes that further elaborated on
Weyl's ideas \cite{Einstein1921}, notwithstanding his critical
evaluation of its physical viability. Thirdly, he had lately
picked up on Eddington's approach of basing the theory on the
affine connection rather than on the metric
\cite{Einstein1923a,Einstein1923b,Einstein1923c}.

In order to evaluate Hilbert's claim, let us take a closer look at
Einstein's work along Eddington's approach, as he had published it
in those papers of 1923 to which Hilbert refers. Following
Eddington,%
\footnote{In this paper, we will not deal with Eddington's own
work but only with Einstein's perception of it.} Einstein had
taken the components of a real, symmetric affine connection
$\Gamma^{\kappa}_{\lambda\mu}$ as the basic quantities of the
theory instead of the metric tensor field $g_{\mu\nu}$ which
provided the dynamical variables in the original theory. From the
symmetric connection he had constructed an asymmetric contracted
curvature tensor,
\begin{equation}
R_{kl} = - \Gamma^{\alpha}_{kl,\alpha}
         + \Gamma^{\alpha}_{k\beta}\Gamma^{\beta}_{l\alpha}
         + \Gamma^{\alpha}_{k\alpha,l}
         - \Gamma^{\alpha}_{kl}\Gamma^{\beta}_{\alpha\beta}.
\end{equation}
Since $R_{kl}dx^kdx^l$ is an invariance of the line element, it
was tempting to split the Ricci tensor into a symmetric part
$g_{\rm kl}$, to be interpreted as a metric tensor associated with
the gravitational field, and an antisymmetric part $\phi_{\rm
kl}$, to be associated with the electromagnetic field tensor.

In a first note presented to the Berlin Academy on 15 February
1923, Einstein observed that Eddington had not yet solved the
problem of finding the necessary equations that would determine
the 40 components of the connection. He therefore set out to
provide just such equations. He postulated a Hamiltonian
principle,
\begin{equation}
\delta\left\{\int\mathcal{H} d\tau\right\} = 0,
\end{equation}
with a Lagrangian that would depend only on the contracted
curvature tensor, $\mathcal{H} = \mathcal{H}(R_{kl})$.\footnote{We
are using Einstein's and Hilbert's notation, both of whom referred
to the Lagrangian as a Hamiltonian function.} More specifically,
he proposed a tentative set of field equations for the affine
connection based on a Lagrangian proportional to the square root
of the determinant of the contracted curvature tensor
\begin{equation}
\mathcal{H} = 2\sqrt{-|R_{kl}|}.
\end{equation}
In his first note, Einstein does not proceed to derive the field
equations explicitly from that Lagrangian. Instead, he does the
variation for a general Lagrangian $\mathcal{H}$ which gives him
\begin{equation}
\mathfrak{s}^{kl}_{;\alpha}
- \frac{1}{2}\delta^{k}_{\alpha}\mathfrak{s}^{l\sigma}_{;\sigma}
- \frac{1}{2}\delta^{l}_{\alpha}\mathfrak{s}^{k\sigma}_{;\sigma}
- \frac{1}{2}\delta^{k}_{\alpha}\mathfrak{f}^{l\sigma}_{,\sigma}
- \frac{1}{2}\delta^{l}_{\alpha}\mathfrak{f}^{k\sigma}_{,\sigma}
=0,
\end{equation}
where $\mathfrak{s}^{kl}$ and $\mathfrak{f}^{kl}$ are defined as
variations of $\mathcal{H}$ with respect to $g_{kl}$ and
$\phi_{kl}$, respectively, i.e.\
\begin{equation}
\delta \mathcal{H} = \mathfrak{s}^{kl}\delta g_{kl}
+ \mathfrak{f}^{kl}\delta \phi_{kl}.
\end{equation}
Solving with respect to $\Gamma^{\lambda}_{\mu\nu}$, he obtains
\begin{equation}
\Gamma^{\alpha}_{kl} = \frac{1}{2}s^{\alpha\beta}
\bigg(s_{k\beta,l} + s_{l,\beta,k} - s_{kl,\beta}\bigg) -
\frac{1}{2}s_{kl}i^{\alpha} + \frac{1}{6}\delta^{\alpha}_{k}i^{l}
+ \frac{1}{6}\delta^{\alpha}_{l}i^{k},
\end{equation}
where
$\mathfrak{i}^l=\sqrt{-|s_{kl}}i^l=\mathfrak{f}^{l\sigma}_{,\sigma}$,
and indices are raised and lowered by means of $s_{kl}$ and
$s^{kl}$ respectively, a fundamental tensor which in turn is
defined via $\mathfrak{s}^{kl} = s^{kl}\sqrt{-|s_{kl}|}$ and
$s_{\alpha i}s^{\beta i} = \delta^{\beta}_{\alpha}$.

Explicit field equations were given by Einstein in a short follow
up note to his paper \cite{Einstein1923b} published on May 15,
1923. In it he briefly recapitulated the basic equations of his
previous note, implicitly introducing a change of notation by
denoting the Ricci tensor as $r_{kl}$, and denoting the Ricci
tensor formed from the fundamental tensor $s^{kl}$ only as
$R_{kl}$. The field equations were now given as the symmetric and
antisymmetric parts of
\begin{equation}
r_{kl} = R_{kl} + \frac{1}{6}
\Big[
\Big(i_{k,l} - i_{l,k}\Big)
+ i_k i_l
\Big].
\end{equation}
These field equations would not hold up for long. Already two
weeks after the publication of the second note, Einstein presented
a third note to the Prussian Academy dealing with the affine
theory \cite{Einstein1923c}, published in the Academy's
Proceedings on 28 June. While Einstein in the introductory
paragraph of that paper announced that ``further considerations''
(``Weiteres Nachdenken'') had led him to a ``perfection''
(``Vervollkommnung'') of the theory laid out in the previous two
notes, he was, in fact, going to present some major revisions,
including a new set of field equations.

One change in his understanding is reflected in an implicit
overall change of notation. While he had previously regarded the
symmetric and antisymmetric parts $g_{kl}$ and $\phi_{kl}$ of the
Ricci tensor $R_{kl}=R_{kl}(\Gamma^{\lambda}_{\mu\nu})$ as the
``metric and electromagnetic field tensors,'' he now attaches this
physical meaning to different quantities. Hence he now denotes the
symmetric part of $R_{\mu\nu}$ as $\gamma_{\mu\nu}$ and uses the
letter $g$ resp.\ $\mathfrak{g}$ to denote the quantities that he
had previously denoted by $s$ resp.\ $\mathfrak{s}$,
\begin{equation}
\delta \mathcal{H} = \mathfrak{g}^{\rm kl}\delta \gamma_{\rm kl}
                  + \mathfrak{f}^{\rm kl}\delta \phi_{\rm kl}.
\label{eq:compdiff}
\end{equation}
It is the quantities $\mathfrak{g}^{\rm kl}$ and
$\mathfrak{f}^{\rm kl}$ that were now ``regarded as tensor
densities of the metric and electric field.'' Einstein also
pointed out that he no longer would assume the Lagrangian
$\mathcal{H}$ to depend on $R_{\mu\nu}$, i.e.\ only on the sum of
$\gamma_{\mu\nu}+\phi_{\mu\nu}$ but would now allow for the
possibility that it depend on $\gamma_{\mu\nu}$ and
$\phi_{\mu\nu}$ independently.

Thirdly, Einstein does not simply proceed to discuss restrictive
conditions or other motivations for a definite choice of
$\mathcal{H}$ in order to fix the field equations. Instead, he
argues that since, by assumption, eq.~(\ref{eq:compdiff}) is a
complete differential,
\begin{equation}
\gamma_{\mu\nu}d\mathfrak{g}^{\mu\nu} +
\phi_{\mu\nu}d\mathfrak{f}^{\mu\nu}
\end{equation}
is a complete differential of another scalar density
$\mathcal{H}^{\ast}$ where $\mathcal{H}^{\ast}$ is a function of
the tensor densities of the metric and electric fields,
$\mathcal{H}^{\ast} =
\mathcal{H}^{\ast}(\mathfrak{g}^{\mu\nu},\mathfrak{f}^{\mu\nu})$.
For the choice of a definite $\mathcal{H}^{\ast}$ Einstein then
gives some arguments. It should be a function of the two
invariants of the electromagnetic fields, and specifically, he
argues that, ``according to our present knowledge, the most
natural ansatz''\footnote{``Der im Sinne unserer bisherigen
Kenntnisse nat\"urlichste Ansatz'' \cite[p.~139]{Einstein1923c}.}
would be
\begin{equation}
\mathcal{H}^{\ast} = 2\alpha\sqrt{-g} -
\frac{\beta}{2}f_{\mu\nu}\mathfrak{f}^{\mu\nu}.
\end{equation}
The resulting field equations, after a rescaling of the
electromagnetic field, read
\begin{align}
R_{\mu\nu} - \alpha g_{\mu\nu}
&=
- \Big[\Big(-f_{\mu\sigma}f_{\nu}^{\sigma}
  + \frac{1}{4}g_{\mu\nu}f_{\sigma\tau}f^{\sigma\tau}\Big)
  + \frac{1}{\beta}i_{\mu}i_{\nu}\Big] \label{eq:fielda} \\
-f_{\mu\nu} &= \frac{1}{\beta}i_{[\mu,\nu]}.
\label{eq:fieldb}
\end{align}
For us, the last half-page of his note, following immediately
after equations (\ref{eq:fielda}), (\ref{eq:fieldb}) is most
interesting. Einstein observed that the field equations derived
along the lines sketched above may also be derived, in fact quite
easily, from a different Hamiltonian principle. He conceived of
$\mathcal{H}$ as a function of $\mathfrak{g}^{\mu\nu}$ and
$\mathfrak{f}^{\mu\nu}$ which Einstein, as was mentioned, in this
third note took to be the tensor densities of the metric and
electromagnetic fields, $\mathcal{H} =
\mathcal{H}(\mathfrak{g}^{\mu\nu},\mathfrak{f}^{\mu\nu})$. The
Lagrangian whose variation with respect to $\mathfrak{g}^{\mu\nu}$
and $\mathfrak{f}^{\mu\nu}$ would produce the field equations
(\ref{eq:fielda}), (\ref{eq:fieldb}) directly then reads
\begin{equation}
\mathcal{H} = \sqrt{-g}\Big[ R-2\alpha
  + \kappa \Big(\frac{1}{2}f_{\sigma\tau}f^{\sigma\tau}
                -\frac{1}{\beta}i_{\sigma}i^{\sigma}\Big)
                \Big].
\end{equation}
Here $R$ denotes the Riemannian curvature scalar formed from the
metric tensor $g_{\mu\nu}$. Notwithstanding the cosmological
constant term $-2\alpha$, the Lagrangian already looks familiar.
But we need one more little step. In the penultimate paragraph of
his paper, Einstein suggests that for a physical interpretation it
would be most useful to introduce the ``electromagnetic
potential''
\begin{equation}
-f_{\mu} = \frac{1}{\beta}i_{\mu},
\label{eq:elmpotential}
\end{equation}
a step that would eventually turn the field equations into those
that were identical - up to the sign of the constant $\beta$ - to
field equations proposed by Weyl.

Let us now pause and look at Einstein's result through Hilbert's
eyes. If we substitute the electromagnetic potential
(\ref{eq:elmpotential}) for $i_{\mu}$, we get the variational
principle in the form
\begin{equation}
\delta \mathcal{H} =
\delta\int\Big\{R-2\alpha + \kappa\big(\frac{1}{2}
f_{\sigma\tau}f^{\sigma\tau}
- \beta f_{\sigma}f^{\sigma}\big)\Big\}\sqrt{-g}d\tau = 0.
\end{equation}
Comparing this variational principle with the variational integral
(\ref{eq:varprinc1}) given by Hilbert in his lectures, we see that
Hilbert's interpretation actually does capture Einstein's result
of his third note on the affine theory, provided we make the
following identifications. Hilbert's $K$ would be Einstein's
$R-2\alpha$, i.e.\ Hilbert ignored the cosmological term. However,
such a term would fit easily into Hilbert's original scheme. We
would also identify Hilbert's $\alpha\Phi$ with Einstein's
$\frac{\kappa}{2} f_{\sigma\tau}f^{\sigma\tau}$. Finally, we would
identify Hilbert's $\beta\varphi$ with Einstein's $\kappa\beta
f_{\sigma}f^{\sigma}$.

One technical difference remains. Hilbert is doing the variation
with respect to the electromagnetic potential $\varphi_{\mu}$
whereas Einstein is doing the variation with respect to the
electromagnetic tensor density $\mathfrak{f}^{\mu\nu}$. In
Hilbert's theory, the electromagnetic field was {\it defined} as
$\Phi_{mn} \equiv \varphi_{[m;n]}$ and the variation {\it
produced} the generalized Maxwell equations (\ref{eq:elm}). In
Einstein's theory, the variation is done with respect to the field
$\mathfrak{f}^{\mu\nu}$. The variation of the term $\beta
f_{\sigma}f^{\sigma}$ makes use of the {\it definition} $f^{\mu} =
-(1/\beta)f^{\mu\nu}_{;\nu}$ and {\it produces} the relation
$f_{\mu\nu} = f_{[\mu;\nu]}$ as an electromagnetic field equation.
Taking into account that for symmetric connections the homogeneous
Maxwell equations (\ref{eq:homMaxwell}) follow from the fields
being given as the rotation of a vector, we can now see the point
of Hilbert's remark.

Regardless of how Einstein had derived his field equations in the
first place, he himself had cast them into a form that was
technically equivalent to Hilbert's initial framework of 1915. The
resulting equations were essentially equivalent to Hilbert's with
the only difference that what appeared as a definition and a field
equation in one framework turned out to be the resulting field
equations and the defining relation in the other. In Hilbert's
words, the difference amounted to an ``interchange of the two
series of Maxwell equations.'' To be sure, the identification
involves some amount of interpretation but essentially we can see
why Hilbert rejoiced:
\begin{quote}
And if on a colossal detour via Levi-Civita, Weyl, Schouten,
Eddington, Einstein returns to this result, then this certainly
provides a beautiful confirmation.\footnote{``Und wenn auf dem
kollossalen Umweg \"uber Levi Civita, Weyl, Schouten, Eddington
Einst. zu dem Resultat zur\"uckgelangt, so liegt darin sicher eine
sch\"one Gew\"ahr.'' {\it Lectures}, part I, p.~20.}
\end{quote}
It also becomes conceivable that Hilbert's reprint of his 1915 and
1917 notes on the {\it Grundlagen der Physik} in 1924 as a single
paper in the {\it Mathematische Annalen} was not motivated by his
desire to revise his original theory (as has been argued in
\cite{RennStachel1999}). His lectures of 1923 in Hamburg and
Zurich rather suggest that the true motivation for Hilbert becomes
visible on the background of his perception of Einstein's latest
work on the affine theory. He saw Einstein's work as a
confirmation of his original approach. Hence, there is no reason
to assume that Hilbert did not believe what he wrote about his
original 1915 theory in the new introduction to the 1924 reprint:
\begin{quote}
I firmly believe that the theory which I develop here contains a
core that will remain and that it creates a framework that leaves
enough room for the future construction of physics along the field
theoretic ideal of unity.\footnote{``Ich glaube sicher, da{\ss}
die hier von mir entwickelte Theorie einen bleibenden Kern
enth\"alt und einen Rahmen schafft, innerhalb dessen f\"ur den
k\"unftigen Aufbau der Physik im Sinne eines feldtheoretischen
Einheitsideals gen\"ugender Spielraum da ist.''
\cite[p.~2]{Hilbert1924}.}
\end{quote}

\section{Accessorial Laws of Nature?}

As we have seen, Hilbert meant what he said, even though he was
deliberately formulating his claim as a hypothesis. Having
established that his ``world equations'' are confirmed, if only by
his own perception of a convergence of related research efforts,
Hilbert in his second lecture became somewhat more speculative. Of
central importance for the argument of his second lecture is the
notion of ``accessorial laws.''\footnote{For another discussion of
this concept, see \cite{MajerSauer2003}.} While Hilbert does use
the term ``accessorial" in a contemporary lecture
course,\footnote{See lecture notes for course on ``\"Uber die
Einheit in der Naturerkenntnis,'' held in winter 1923/24. NSUB
Cod.~Ms.\ Hilbert 568, p.~247.} we are not aware of any other
usage of the term, neither in Hilbert's own Oeuvre nor in any of
his contemporaries' writings. Our guess is that Hilbert created a
neologism based on the Latin ``accedere"
--- in its meaning ``to add.''\footnote{We realize that the English
word ``accessorial'' is not a neologism and its meaning of
auxiliary, supplementary makes good sense in the present context.}
What notion then does Hilbert want to capture by the term
``accessorial''? He says:
\begin{quote}
Anything that needs to be added to the world equations in order to
understand the events (``Geschehnisse") of inanimate nature, I
will briefly call ``accessorial."\footnote{``Ich m\"ochte Alles,
was noch zu den Weltgleichungen hinzugef\"ugt werden muss, um die
Geschehnisse in der leblosen Natur zu verstehen, kurz {\it
accessorisch} nennen." {\it Lectures}, part II, p.~1.}
\end{quote}
An obvious candidate for something ``accessorial" with respect to
the ``world equations" immediately comes to mind. These equations
being differential equations, require for the explanation of
``events" certainly the determination of initial or boundary
conditions. Indeed, Hilbert concedes that initial or boundary
conditions are necessary in order to allow for definite solutions
of the ``world equation,''\footnote{For further discussion Hilbert
would assume the world to be Euclidean-Newtonian at infinity, but
with respect to contemporary cosmological debates Hilbert added a
disclaimer to the effect that this choice was only motivated by
formal simplicity and was made only to fix the ideas.} but,
obviously he has something more demanding than ``initial
conditions'' in mind, because he does not qualify them as
``accessorial.'' Hence, the question arises, what else does he
want to capture with the term ``accessorial.'' To answer this
question, we have to explain how he proceeds in the second
lecture.

Conceding that the world equations are in need of initial or
boundary conditions, the main point of Hilbert's second Hamburg
lecture is to argue for another and non-trivial meaning of
``accessorial.'' Even with initial conditions, the equations,
being differential equations with respect to some time coordinate,
would only predict the future from the past, but would they also
teach us something about the present which after all, as Hilbert
argues, is what we really want? If the answer is no, then we are
in need of ``accessorial'' {\it laws}, that can tell us something
about the present state of nature. Now the interesting point is,
as we will see in a moment, that Hilbert argues that no such
accessorial laws of nature exist, for the simple reason that
precisely that which we want to capture with such laws is either
inconsistent with the world equations or already contained in
them.

A first argument supporting his claim is a discussion of the
irreversibility of thermodynamics. He looks at the example of the
mixing of a gas that is initially distributed over two separate
halves of a container and emphasizes that the apparent asymmetry
with respect to past and future is exclusively a consequence of
the choice of the initial states and the initial conditions, and
hence that the irreversibility is not one that exists objectively
in inanimate nature and its lawfulness but is only an apparent
irreversibility, arising from what he called our anthropomorphic
point of view.

The argument is interesting in itself, especially with respect to
Hilbert's epistemological position.\footnote{For a more detailed
discussion of the non-objective but anthropomorphic character of
certain apparently irreversible processes in inanimate nature, see
\cite{Majer2002}.} While Hilbert is unambiguous about his claim
that there are no accessorial laws introduced in statistical
mechanics, he himself brings up an obvious objection. The example
of the diffusion of a gas in a container in the theoretical
context of kinetic gas theory presupposes the assumption that
there exist atoms and molecules, and that these are the
fundamental constituents of the diffusing gas. This argument leads
him to a discussion of the question whether the principle of
atomism is an accessorial law of nature. Hilbert's position on
this issue is just as unambiguous as is his position on the issue
of irreversibility. He argues that the world equations, possibly
after necessary elaborations or corrections, suffice to explain
the existence, and even the structure, and properties, of matter.
In order to justify this claim, Hilbert refers to Bohr's quantum
theory and to the explanation of basic features of the periodic
system of elements (such as its periodicity and the chemical
stability of the inert gases) on the grounds of the electron orbit
model.

Hilbert's conclusion from this discussion is that the field
equations and laws of motion suffice to derive the deepest
properties of matter including the characteristic details of the
chemical elements as particular mathematical integrations of the
field equations. It is important to note that in this respect the
``world equations'' differ fundamentally from Newton's laws,
including gravity, because the latter do not imply anything about
the existence of atoms and molecules. Of course, Hilbert would
take it for granted, among other things, that particle-like
solutions of the field equations would exist whose dynamics would
then be governed by the field equations as well, rather than by
independent equations of motion.

Hilbert's belief that the world equations can tell us something
about the present presupposes that we accept only those solutions
to the equations that correspond to constant or periodic processes
in nature. Hence we have to qualify the assertion about the
non-existence of accessorial laws by admitting that there are
accessorial {\it ideas} and {\it principles}, such as stability
and periodicity. But the crucial difference, according to Hilbert,
is that these accessorial ideas and principles do not have the
character of new equations but are of a more general nature that
is connected to our thinking as such and to our attitude towards
nature.

It so happened that a number of the assumptions made by Hilbert,
both explicitly and implicitly, turned out to be highly
problematic, if not false. This is the case, e.g, with the
violation of gauge invariance implied by accepting an explicit
dependence of the Lagrangian on the electromagnetic potential. But
before dismissing Hilbert as a bad speculative physicist, let us
take seriously the fact that he himself in a most enthusiastic
manner pointed to the rapid development of the natural sciences
and the rapid succession of fundamental discoveries. It seems to
us that his perhaps premature acceptance of results which had yet
to be confirmed appears to us today na\"{\i}ve for a very specific
reason. Hilbert's optimism was fuelled by his unwillingness to
accept the fact that the modern development of the natural
sciences no longer allows for a conceptual unity of knowledge. In
this respect, by the way, he was not alone. Indeed, the purpose of
the first two lectures of his trilogy was to provide the
scientific underpinning for a more philosophical claim that he
made in the third lecture.

\section{Hilbert's Position between Kantian Apriorism and
Poincar\'e's Conventionalism}

Let us therefore return now to Hilbert's epistemological
position.\footnote{See \cite{MajerSauer2003} for a more extensive
discussion.} In his third lecture, Hilbert addresses the ancient
question as to the sources of our knowledge, or, in his own words:
\begin{quote}
We are dealing here with a decision of an important philosophical
problem, namely the old question as to the portion of our
knowledge that comes from our thinking, on the one hand, and from
experience, on the other hand.\footnote{``Wir stehen da vor der
Entscheidung \"uber ein wichtiges philosophisches Problem,
n\"amlich vor der alten Frage nach dem Anteil, den das Denken
einerseits und die Erfahrung andererseits an unserer Erkenntnis
haben." {\it Lectures}, part III, p.~1.}
\end{quote}
In the remainder of this paper we want to say a few words about
Hilbert's answer to the question of the borderline between
knowledge a priori and knowledge by experience. Hilbert's position
is based on two fundamental presuppositions. The first of these is
the distinction between two different domains of the natural
sciences, the domain of ``inanimate" nature, which is the proper
domain of physics in the widest sense, and the domain of living
beings including ``man as such" which is the domain of biology,
including the social and cultural sciences. Even though the
distinction seems problematic from a physicalistic point of view,
it has not been shown to this day whether the laws of physics, as
we know them today, suffice to deduce the phenomena of life, or
whether we need in fact some accessorial laws or
principles.\footnote{For a discussion of this intricate question
in connection with the supposed irreversibility of living
processes, see \cite{Majer2002}.} But for our context, it is
sufficient to take this distinction as a warning that the claim
that there exist ``world equations" in the strong sense, i.e.\
that we do not need any accessorial laws, is certainly more
difficult to establish if the life sciences were included in the
claim.

The second fundamental distinction that plays a role here is a
distinction between three different levels of experience: (i) a
level of every day experience, (ii) a level of scientific
experience in the broadest sense of the term, and (iii) a level of
totally objective knowledge that is achieved by an emancipation
from what Hilbert calls our anthropomorphic point of view. The
principle of objectivity that Hilbert had introduced earlier in
his first lecture illustrates what Hilbert has in mind by the
emancipation from the anthropomorphic point. This principle states
\begin{quote}
A sentence about nature, expressed in coordinates, is only then a
proposition about the objects in nature, if the sentence has a
content which is independent of the coordinates.\footnote{``Ein in
Koordinaten ausgedr\"uckter Satz \"uber die Natur ist nur dann
eine Aussage \"uber die Gegenst\"ande in der Natur wenn er von den
Koordinaten unabh\"angig einen Inhalt hat.'' {\it Lectures},
part~I, p.~3.}
\end{quote}
According to Hilbert, this emancipation from the coordinate system
can be achieved in three different ways that correspond to the
three forms of singular, particular, and general judgment: First,
by showing or presenting a concrete object, in respect to which
the coordinate system has to be fixed; second, in the form of an
existential assertion by saying: there exists a coordinate system
in which all the formulated relations between the objects
considered are valid; third, by formulating the proposition in a
form that is valid in every coordinate system. Evidently, this
distinction implies that the introduction of coordinates in the
first place is a compromise to our human way of looking at nature,
and the third way of emancipating from a coordinate system
therefore represents the most far-reaching ``emancipation from the
anthropomorphic point of view.'' A certain view of the actual and
proper development of science is implicit in this latter
assumption, and Hilbert's epistemology is indeed a philosophy of
graded progress \cite{MajerSauer2003}.

Hilbert emphatically points out that the Kantian question is ripe
for an answer for two reasons, (1) the spectacular progress in the
sciences of the time and (2) the advent of the axiomatic method.
Much more needs to be said about Hilbert's epistemological
position in general and the interrelations of these two moments in
particular. But for the sake of brevity, let us here only point to
the role of the world equations. Hilbert says:
\begin{quote}
If now these world equations, and with them the framework of
concepts, would be complete, and we would know that it fits in its
totality with reality, then in fact one needs only thinking and
conceptual deduction in order to acquire all physical
knowledge.\footnote{``Wenn nun diese Weltgleichungen und damit das
Fachwerk vollst\"andig vorl\"age, und wir w\"ussten, {\it dass es}
auf die Wirklichkeit in ihrer Gesamtheit passt und dann bedarf es
tats\"achlich nur des {\it Denkens} d.h.\ der begrifflichen {\it
Deduktion}, um alles phys.\ Wissen zu gewinnen.'' {\it Lectures},
part~III, pp.~20f. (Hilbert's emphasis).}
\end{quote}
Leaving aside the difficult question concerning completeness of
physical theories, we only wish to emphasize that Hilbert,
contrary to what one might expect from this quote, by no means
wants to take an idealistic position. He emphasizes
\begin{quote}
I claim that precisely the world equations can be obtained in no
other way than from experience. It may be that in the construction
of the framework of physical concepts manifold speculative view
points play a role; but whether the proposed axioms and the
logical framework erected from them is valid, experience alone can
decide this question.\footnote{``... behaupte ich, dass gerade die
Weltgesetze auf keine andere Weise zu gewinnen sind, als aus der
Erahrung. M\"ogen bei der Konstruktion des Fachwerkes der phys.
[Begriffe] mannigfache spekulative Gesichtspunkte mitwirken: {\it
ob} die aufgestellten Axiome und das aus ihnen aufgebaute logische
Fachwerk stimmt, das zu {\it entscheiden}, ist allein die {\it
Erfahrung} im Stande.'' ibid., p.~21 (Hilbert's emphasis).}
\end{quote}
In the sequel to the lecture, Hilbert refined this somewhat crude
position by taking issue with Kantian apriorism and with
Poincar\'e's conventionalism. The upshot is
\begin{quote}
The opinion advocated by us rejects the absolute Apriorism and the
Conventionalism; but nevertheless it does in no way retreat from
the question of the precise validity of the laws of nature. I will
instead answer this question in the affirmative in the following
sense. The individual laws of nature are constituent parts of the
total conceptual framework, set up axiomatically from the
world-equations. The world-equations are the precipitation of a
long, in part very strenuous, experimental inquiry and of
experience, often delayed by going astray. In this way we come to
the idea that we approximate asymptotically a Definitivum by
continued elaboration and completion of the
world-equations.\footnote{``Die von uns vertretene Meinung
verwirft den unbedingten Apriorismus und den Konventionalismus;
aber sie entzieht sich trotzdem keineswegs der vorhin
aufgeworfenen Frage nach der genauen G\"ultigkeit der
Naturgesetze. Ich m\"ochte diese Frage vielmehr bejahen und zwar
in folgendem Sinne. Die einzelnen Naturgesetze sind Bestandteile
des Gesamtfachwerkes, das sich  aus den Weltgleichungen
axiomatisch aufbaut. Und die Weltgleichungen sind der Niederschlag
einer langen zum Teil sehr m\"uhsamen und oft durch Irrwege
aufgehaltenen experimentellen Forschung und Erfahrung. Wir
gelangen dabei zu der Vorstellung, daß wir uns durch fortgesetzte
Ausgestaltung und Vervollst\"andigung der Weltgleichungen
asymptotisch einem Definitivum n\"ahern.'' ibid., pp.~42f.}
\end{quote}
Whatever may be said about this position from a historical and
philosophical standpoint, we hope to have at least shown that
Hilbert's work along the unified field theory program is embedded
in a broader perspective of epistemological and methodological
concerns that well deserves to be taken seriously, even on today's
philosophical horizon.

\section*{Acknowledgments}

One of us (T.S.) would like to thank Jim Ritter for illuminating
discussions about Einstein and the unified field theory program
during our common stay at the Max Planck Institute for the History
of Science in Berlin. We also thank Diana Buchwald, Dan Kennefick,
and Stephen Speicher for helpful comments on an earlier draft of
this paper. Hilbert's lectures are quoted by kind permission of
the {\it Nieders\"achsische Staats- und Universit\"atsbibliothek
(Handschriftenabteilung)}.

%
%
%

\end{document}